\documentclass[pre,twocolumn]{revtex4-1}

\usepackage{graphicx}

\usepackage{amssymb,amsfonts,amsmath}
\usepackage[usenames,dvipsnames]{color}

    \def \a{\alpha}
\def \w{\omega}

\def \>{\rangle} 
\def \<{\langle} 
 
\def\be{\begin{equation}} 
\def\ee{\end{equation}} 
\def\longrightharpoonup{\relbar\joinrel\rightharpoonup}
\def\longleftharpoondown{\leftharpoondown\joinrel\relbar}

\def\longrightleftharpoons{
  \mathop{
    \vcenter{
      \hbox{
      \ooalign{
        \raise1pt\hbox{$\longrightharpoonup\joinrel$}\crcr
	  \lower1pt\hbox{$\longleftharpoondown\joinrel$}
	  }
      }
    }
  }
}

\newcommand \bea {\begin{eqnarray}} 
\newcommand \eea {\end{eqnarray}}

\begin{document}

\title{ Kuramoto model with coupling through an external medium}

\author{David J. Schwab}\thanks{Equal contribution}
\affiliation{Dept. of Molecular Biology and Physics, Princeton University, Princeton, NJ 08854}
\author{Gabriel G. Plunk}\thanks{Equal contribution}
\affiliation{Princeton Plasma Physics Laboratory, Princeton University, Princeton, NJ 08854}
\author{Pankaj Mehta}
\affiliation{Dept. of Physics, Boston University, Boston, MA 02215}

\begin{abstract}
Synchronization of coupled oscillators is often described using the Kuramoto model. Here we study a generalization of the Kuramoto model where oscillators communicate with each other through an external medium. This generalized model exhibits interesting new phenomena such as bistability between synchornization and incoherence and a qualitatively new form of synchronization where the external medium exhibits small-amplitude oscillations.  We conclude by discussing the relationship of the model to other variations of the Kuramoto model including the Kuramoto model with a bimodal frequency distribution and the Millennium Bridge problem.

\end{abstract}

\maketitle

\section{Introduction}

Spontaneous synchronization of coupled oscillators occurs in many biological, physical, and social systems \cite{strogatz2003sync, mehta2010approaching, schwab2010rhythmogenic}. The onset of synchronization is often described using the Kuramoto model \cite{strogatz2000kuramoto} where oscillators are directly coupled to each other through their phase differences. However, in many realistic systems, the coupling between oscillators is not direct and instead occurs through a shared external medium. Examples include populations of synthetically-engineered bacteria \cite{danino2010synchronized} and single-cell eukaryotes communicating through chemical signals \cite{gregor2010onset}, lasers coupled through a central hub \cite{zamora2010crowd}, and even pedestrians walking on a bridge \cite{gc2005crowd, eckhardt2007modeling}. In contrast to directly coupled oscillators, such systems have received relatively little attention \cite{russo2010global, schwab2010dynamical}. 

In this paper, we present a generalization of the Kuaramoto model where oscillators communicate through a common external medium. As in the usual Kuramoto model, each oscillator is described by a time-dependent phase, $\theta_i(t)$, which in the absence of coupling, rotates at its natural frequency $\omega_i$. We concentrate on the case when the number of oscillators is large, $N \gg 1$ and the natural frequencies are assumed to be drawn from a prescribed distribution function $g(\omega)$ with mean frequency $\omega_0$. The oscillators are coupled through an external medium which has an amplitude and phase and is described using a complex number $Z(t)=R(t)e^{i \Phi(t)}$. The model can be derived from the system introduced in \cite{schwab2010dynamical} to study dynamical quorum sensing by considering the weak interaction limit where the amplitude of each individual oscillator's limit cycle remains approximately constant.

The dynamics of the external medium gives rise to interesting new phenomena not seen in directly coupled oscillators. A key dynamical parameter is the density, $\rho$, of oscillators.  For large densities, $\rho \rightarrow \infty$, our model reduces to the usual Kuramoto model. Surprisingly, in the limit $\rho \rightarrow 0$, the phase diagram of our model can be mapped to the phase diagram of a Kuramoto model with a bimodal frequency distribution \cite{martens2009exact} with the added restriction that the dynamics is constrained to lie within the Ott-Antonsen manifold \cite{ott2008low}.  At low but non-zero densities, $\rho \ll 1$, the system exhibits bistability between incoherence and synchronization as well as between two difference kinds of synchronized states.  Additionally, when $\omega_0=0$, the dynamics of our model is in one-to-one correspondence with that of the Millennium Bridge problem. Thus, the Kuramoto model with coupling through an external medium represents a particularly simple physical model which captures behavior exhibited by a large number of other Kuramoto-like models .

The outline of the paper is as follows. In section \ref{sec:redphase}, we define our model and show how it naturally arises from considering the weak-coupling limit of the dynamic quorum sensing model considered in \cite{schwab2010dynamical}. In section \ref{sec:dynamics}, we use the Ott-Antonsen ansatz to derive equations for the dynamics of our model for an arbitrary oscillator-frequency distribution. In the next section \ref{sec:lordynamics}, we  consider the special case of  Lorentzian distribution where we can derive analytic equations for the stability of various phases. In section \ref{sec:phasediagram}, we use these equations and numerical simulations to construct a phase diagram for our model. In section \ref{sec:relmodel}, we discuss the relationship of our model to other problems including the Kuramoto model with bimodal frequency distribution and the Millennium Bridge problem. We then discuss the implications of model and conclude.



\section{Derivation of the model}
\label{sec:redphase}
The Kuramoto model with oscillators coupled through a shared external medium can be derived by considering the weak coupling limit of the model introduced in \cite{schwab2010dynamical}. The model in (DQS) consists of a population of limit-cycle oscillators, $z_i$, each with a natural frequency $\omega_j$, diffusively coupled to a shared medium, $Z$. When chemicals leave the oscillators and enter the medium, they are diluted by a factor $\alpha=V_{int}/V_{ext} \ll 1$, the ratio of the volume of the entire system to the that of an individual oscillator. Furthermore, the external media is degraded at a rate $J$. The dynamics of the system are captured by the equation
\bea
\frac{dz_j}{dt}&=& (\lambda_0+i \omega_j - |z_j|^2)z_j -D(z_j-Z)\label{zjeq} \\
\frac{dZ}{dt} &=& \alpha D \sum_j( z_j -Z) -J Z \label{Zeq1}
\eea
with  $\omega_j$ drawn from a distribution $h(\omega)$ which we assume to be an even function about a mean frequency $\omega_0$.  By introducing a dimensionless density, $\rho= \alpha N$, and shifting to a frame rotating with frequency $\omega_0$, we can rewrite (\ref{Zeq1}) above as
 \be
\frac{dZ}{dt} = \frac{\rho D}{N} \sum_j( z_j -Z) -(J+i\omega_0) Z, \label{Zeq2}
 \ee
 where the frequencies $\omega_j$ are now drawn from a distribution $g(\omega)$ with mean zero.
 
 We are interested in the limit $\lambda_0\gg D$ where individual oscillators are weakly coupled. In this limit, , the amplitude of the  limit cycle is not modified by the coupling and the dynamics is well-described by modeling each oscillator by a single phase variable  \cite{kuramoto2003chemical}.  Explicitly, rewriting  (\ref{zjeq}) in polar coordinates with $z_i=r_ie^{i\theta_i}$ and $Z=Re^{i\Phi}$ yields
 \bea
\frac{dr_j}{dt} &=& (\lambda_0-D-r_j^2)r_j+ DR\cos{(\Phi-\theta_j)}  \label{rjeq}\\ 
\frac{d\theta_j}{dt} &=& \omega_j +\frac{DR}{r_j} \sin{(\Phi-\theta_j)}. \label{PE1}
\eea
 In the large $\lambda_0$ limit, the first term dominates the right-hand side of (\ref{rjeq}) and thus $r_i\simeq\lambda_0^{1/2}$ in steady-state. Defining $\tilde{r_i}=r_i/\lambda_0^{1/2}\simeq1$ and $\tilde{R}=R/\lambda_0^{1/2}$, we are left with the reduced equations that define our model:
 \bea
\frac{d\theta_j}{dt} &=& \omega_j +D\tilde{R} \sin{(\Phi-\theta_j)} \label{PE2} \label{Zeq1a} \\
 \frac{d\tilde{Z}}{dt} &=& \frac{\rho D}{N} \sum_j( e^{i \theta_j} -\tilde{Z}) -(J+i\omega_0) \tilde{Z} \label{Zeq1b}
 \eea
 We will drop the tilde in the remainder of this paper and note that $r=1$ will correspond to the fully synchronized state. Interestingly, however, we will find that 
 
Finally, it is useful to define the Kuramoto order parameter,
 \be
 \bar{z} = \frac{1}{N} \sum_j e^{i \theta_j}.
 \ee
$\bar{z}$ is one when all the oscillators have the same phase and zero when they are completely out of phase. Notice that the external medium communicates with the oscillators only through $\bar{z}=re^{i \phi}$ and we can rewrite (\ref{Zeq1b}) in terms of the order parameter as
\be
 \frac{d{Z}}{dt} = \rho D (\bar{z} -{Z})  -(J+i\omega_0) {Z} 
\label{Zeq}
\ee

\section{Dynamical equations for arbitrary frequency distributions}
\label{sec:dynamics}

\subsection{ Thermodynamic limit and the Ott-Antonsen Ansatz}
In this and the next section, we derive equations for the time-dependence of the distribution function as well as the steady-state values of the order parameter within the low-dimensional manifold of states introduced by Ott and Antonsen \cite{ott2008low}. We have explicitly checked that the dynamics of our model is captured by  the Ott-Antonsen ansatz  using numerical simulations. In what follows, we restrict ourselves to the thermodynamic limit, $N \rightarrow \infty$. In this case, the dynamics is well described by the time-dependent  density function $f( \theta, \omega, t)$, with the fraction of oscillators of frequency $\omega$ lying between $\theta$ and $\theta+d\theta$  given by $f d\theta$.

The evolution for $f$  in time is governed by the continuity equation
\be
\frac{\partial f}{\partial t} + \partial_\theta (f \dot{\theta}) =0
\label{continuityeq}
\ee
where dot signifies a derivative with respect to time. Plugging in (\ref{PE2}) and rewriting the result in complex notation gives
\be
\frac{\partial f}{\partial t}+\partial_\theta[f \omega +\frac{D}{4 \pi i}(Ze^{-i\theta}-Z^*e^{i \theta}) f]=0
\ee
Note that unlike the ordinary Kuramoto model \cite{strogatz2000kuramoto}, the external medium, $Z$, enters  in the second term in place of the continuum limit of the usual order parameter, $\bar{z}=\int f e^{i \theta}$. Within the  Ott-Antonsen ansatz, the dynamics lies on a submanifold  where the distribution functions are of the form,
\be
f(\theta,\omega,t)=\frac{g(\omega)}{2\pi}\left(1+\sum_{n=1}^{\infty}\left[ \alpha^ne^{i n \theta}+\mbox{c.c.}\right] \right)
\label{OAA}
\ee
where $\alpha(\omega, t)$ is a function of $\omega$ and $t$,  $g(\omega)$ is the frequency distribution of the oscillator population, and c.c. denotes complex conjugate. Plugging this ansatz into the continuity equation gives
\be
\frac{\partial \alpha}{\partial t}+i\omega \alpha+\frac{D}{2}(Z\alpha^2-Z^*)=0
\label{alphaeq}
\ee

The Ott-Antonsen ansatz can also be used to derive a simple expression for the order parameter $\bar{z}$. In the thermodynamic limit, we can write the order parameter as
\be
\bar{z}(t)= \int d\omega d\theta f(\theta, \omega, t) e^{i \theta}
\ee
Plugging in (\ref{OAA}) and noting that only the term proportional to $e^{-i \theta}$ is non-zero yields the relation
\be
\bar{z}=\int d\w g(\w)\a^*.
\label{OPalpha}
\ee
We will use this definition extensively in what follows.

\subsection{Stability of the incoherent state}

We start by examining the stability of the incoherent state. The calculation is a straightforward generalization of that performed on the original Kuramoto model \cite{matthews1991dynamics}. In the incoherent state, the phase of the oscillators is uniform and $f= \frac{1}{2\pi}.$  Notice that corresponds to choosing  $\alpha(\omega, t)=0$ in (\ref{OAA}). This choice satisfies the dynamical equations (\ref{alphaeq}), (\ref{Zeq}), and (\ref{OPalpha}) when $Z= \bar{z}=0$. To calculate the stability of this state, we  linearize the dynamical equations about the incoherent state by substituting $f=1/2\pi + \epsilon f_1$ into (\ref{continuityeq}) and keep terms linear in $\epsilon$. This is equivalent to keeping only terms linear in $\alpha$ in (\ref{alphaeq}):
\be
\frac{\partial \alpha}{\partial t}+i\omega \alpha-\frac{D}{2}Z^*=0.
\label{linalphaeq}
\ee

For stability analysis, it is sufficient to consider exponential solutions of the form $\alpha=A \exp[(\lambda+i\Omega)t]$ and $Z=Z_0 \exp[(\lambda-i\Omega)t]$, with $\lambda$ and $\Omega$ real numbers. The parameter $\lambda$ determines the stability of the incoherent state and $\Omega$ is a rotation frequency with $\Omega> 0$. In particular, in the lab frame, the steady-state solutions rotate with a frequency $\omega_0-\Omega$, slower than the mean frequency $\omega_0$ due to an external medium induced ``drag". This is in contrast with the usual Kuramoto model with direct coupling where $\Omega=0$. Plugging in solutions of this form into (\ref{linalphaeq}) yields
\be
A=\frac{DZ_0^*}{2(\lambda+i\Omega+i\omega).
}\label{Beqn}
\ee
Using this in eqns. (\ref{OPalpha}) and (\ref{Zeq}) results in a complex equation for $\Omega$ and $\lambda$ of the form
\be
\lambda+i(\omega_0-\Omega)+\rho D+J=\frac{\rho D^2}{2}\int_{-\infty}^\infty \frac{d\omega g(\omega)}{\lambda-i(\omega+\Omega)}
\label{inccomp}
\ee

To determine the boundary of stability, we put $\lambda=0^+$ in the equation above and equate the real and imaginary parts to get
\bea
\frac{\rho D+ J}{\rho D^2} &=& \pi g(\Omega)\label{inc1} \\
\frac{\Omega+\omega_0}{\rho D^2}&=&P \left[ \int_{-\infty}^\infty d\omega \, \frac{g(\omega)}{\omega+\Omega}\right].
 \label{inc2}
\eea
In arriving at this equation, we have used the standard identity $\frac{1}{x+i\epsilon}=P(\frac{1}{x})-i\pi \delta(x)$ when $\epsilon =0^+$, with $P$ denoting principal value. Equation (\ref{inc2}) gives implicit equations for the  surface in the $\rho$-$D$-$J$ parameter space where the incoherence state changes stability. We will use this result to construct the phase diagram for our model below.

\subsection{General Locked Solution}

For general $g(\omega)$, we derive an equation governing locked states as follows. For such solutions, we expect that the order parameter and external medium rotate at a constant frequency $-\Omega$, or equivalently $\omega_0-\Omega$ in the lab frame. We look for solutions of the form
 $\alpha=A(\omega)e^{i\Omega t}, Z=Z_0e^{-i\Omega t}$, and $\bar{z}=\bar{z}_0e^{-i\Omega t}$.
 Plugging these functional forms  into Eq. (\ref{alphaeq}) gives
\be
iA(\omega+\Omega)+\frac{D}{2}(Z_0A^2-Z_0^*)=0\label{appendixref}
\ee
Defining $b=\omega+\Omega$ and $R=|Z_0|$ we see that
\be
A(b)=\frac{-ib\pm\sqrt{-b^2+D^2 R^2}}{DZ_0}.\label{Aeq}
\ee

\begin{figure}[t]
\includegraphics[width=5.0cm]{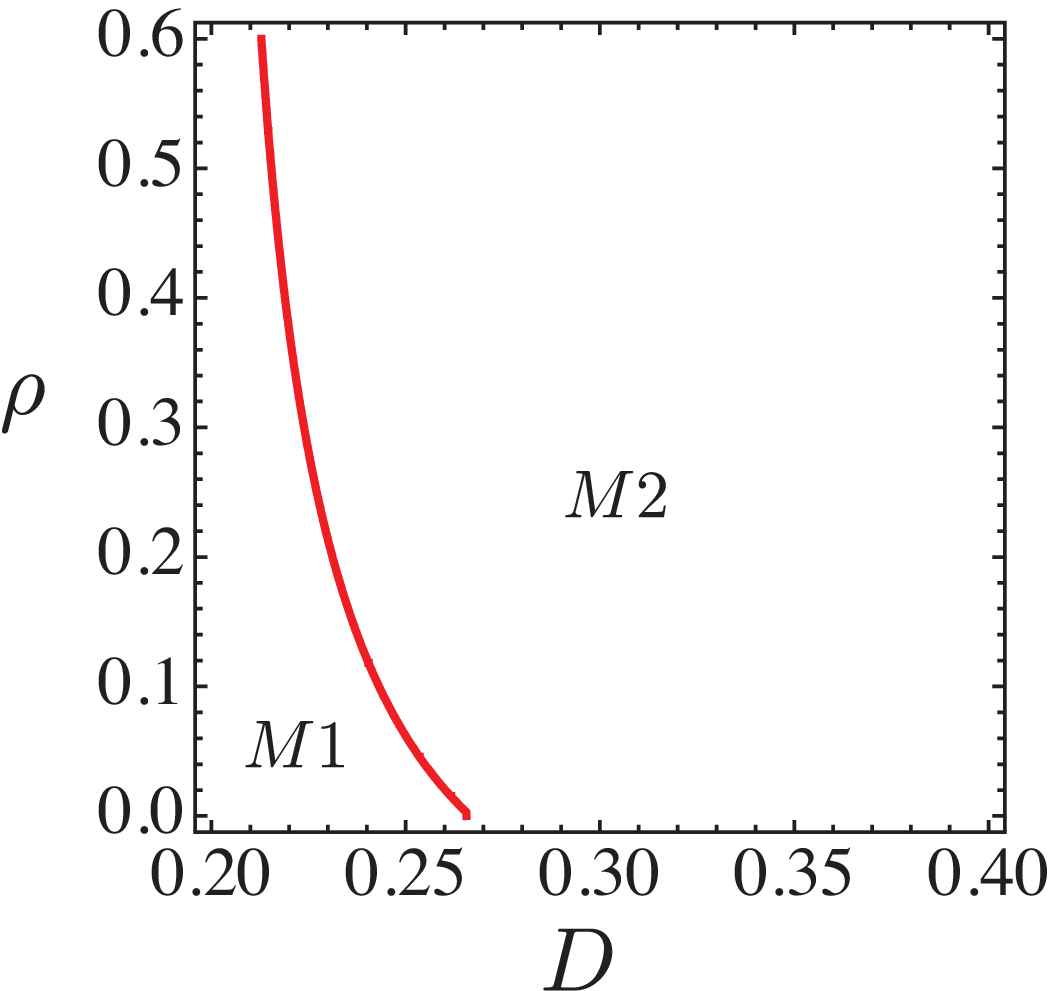}\\
\vspace{0.2cm}
\includegraphics[width=5.0cm]{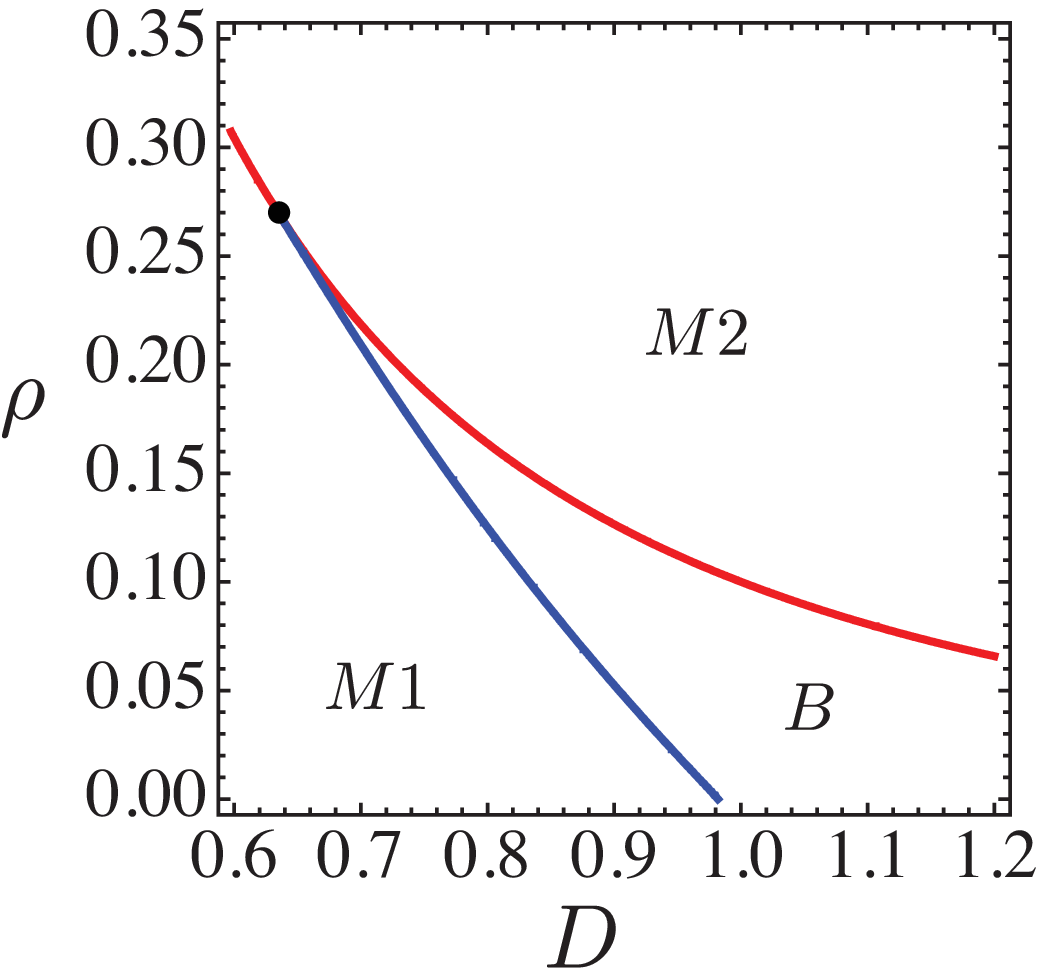}\\
\vspace{0.2cm}
\includegraphics[width=5.0cm]{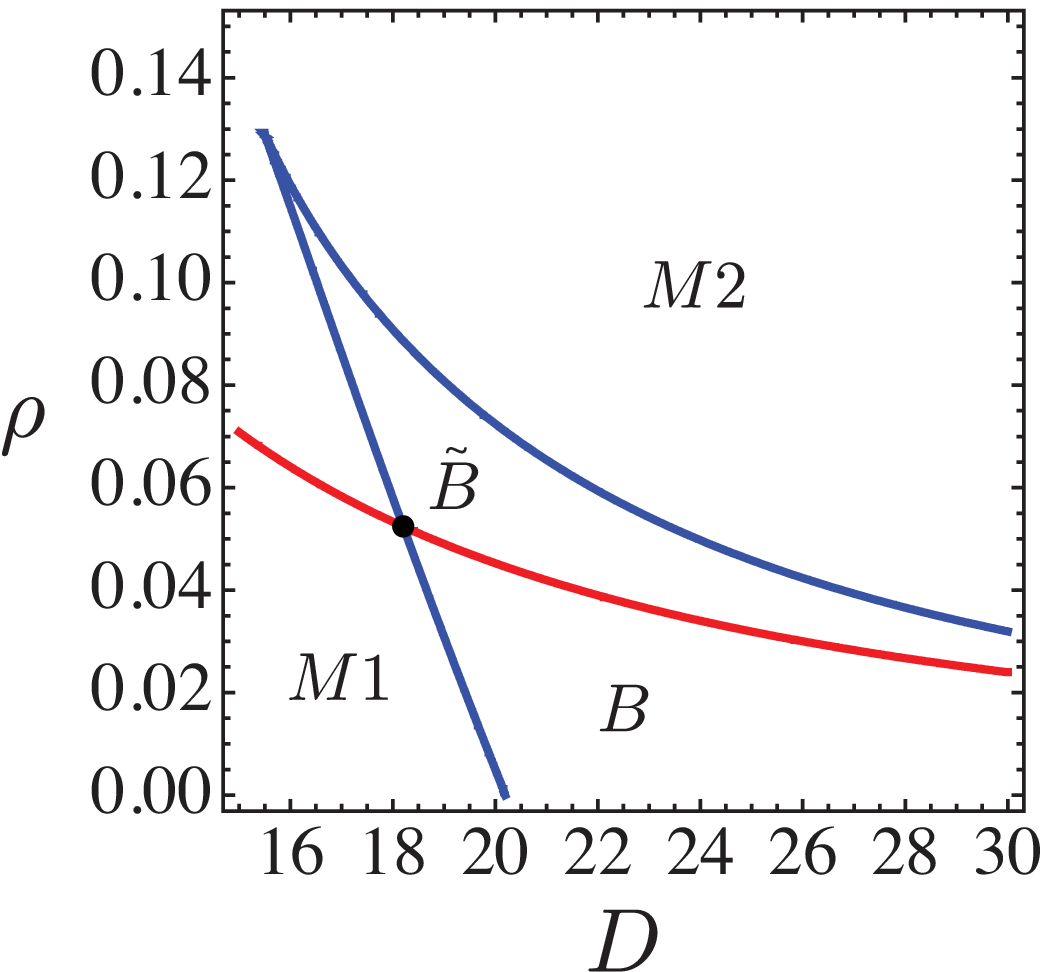}
\caption{Phase diagram as a function of parameters.  $\Delta=0.1$ in all figures. (top) $\omega_0=0.0577$. (middle) $\omega_0=0.4$ (bottom) $\omega_0=10$. $M$ stands for monostability and $B$ stands for bistability. $M1$ - incoherence, $M2$ - coherent oscillations, B - bistability between incoherence and coherent oscillations, and $\tilde{B}$ - bistability between large amplitude phase-locked solution and small amplitude phase-locked solution. Note the change of scale and location of both the horizontal and vertical axes.}
 \label{fig1}
 \end{figure}

To ensure the dynamics stays on the Ott-Antonsen manifold, its necessary to make sure that  $|A(\omega)|\leq 1$ for all $\omega$. In particular, this inequality ensures that the sum in (\ref{OAA}) does not diverge.  This can be guaranteed by properly choosing the positive or negative solutions to (\ref{Aeq}) as follows. Defining 
\be
A \equiv \chi/DZ_0,
\label{defchi}
\ee
 it suffices to choose
\be
\chi(b) =
\begin{cases}
-ib+ \sqrt{-b^2+D^2 R^2}& \text{$b> DR$} \\
-ib \pm \sqrt{-b^2+D^2R^2}& \text{$-DR<b < DR$} \\
-ib- \sqrt{-b^2+D^2R^2}& \text{$b <- DR$}
\end{cases}
\label{chieq}
\ee
Finally, we can plug in this ansatz into (\ref{Zeq}) to get 
\be
[i(\omega_0-\Omega)+\rho D+J]R^2= DZ_0\rho\int db' g(b'-\Omega)A(b')
\label{generaleq}
\ee
Together, (\ref{defchi}), (\ref{chieq}), and (\ref{generaleq}) give a dispersion relationship between $\Omega$ and $R$ for locked solutions.

\section{The Lorentzian Distribution}
\label{sec:lordynamics}

\subsection{Dynamical equations}
In general, the general equations (\ref{generaleq}) derived in the last section must be solved numerically to determine the existence of locked solution.  However, just as in the ordinary Kuramoto model, the mathematic simplifies when $g(\omega)$ is a Lorentzian distribution
\be
g(\omega)=\frac{1}{\pi}\frac{\Delta}{\omega^2+\Delta^2}.
\label{defLor}
\ee 
For Lorentzians, the integrals over $g(\omega)$ appearing in eqns. (\ref{inc1}), (\ref{inc2}), and (\ref{generaleq}) can be performed explicitly and one is left with algebraic equations  These algebraic equations can then be solved to yield the phase boundary of the incoherent state and the existence of locked solutions parameterized by the frequency and amplitude of the external medium, $(\Omega,R)$. One subtlety that remains is that we are not able to analytically determine the stability of locked solutions (except for the incoherent solution), and we therefore resort to numerical solutions to test stability.

The incoherence boundary can be computed explicitly by plugging in (\ref{defLor}) into (\ref{inccomp}) and using contour integration. This yields,
\be
\frac{\Delta \omega_0^2}{\Delta+\rho D + J}\left(1-\frac{\Delta}{\Delta+\rho D + J}\right)+\Delta(\rho D+J)-\rho D^2/2=0
\label{incolor}
\ee
The dynamical equations for locked solutions can also be calculated explicitly for a Lorentzian distribution. The key simplification is that the integral in (\ref{OPalpha}) can again be performed by contour integration so that
 \be
 \bar{z}(t)=\a^*(-i\Delta,t).
 \ee
Thus, evaluating eq. (\ref{alphaeq}) at $\omega = - i \Delta$, we obtain the following equation for the dynamics of the order parameter
\be
\frac{\partial \bar{z}}{\partial t}+\Delta \bar{z}+\frac{D}{2}(Z^*\bar{z}^2-Z)=0.
\label{zbarott}.
\ee
When combined with the equation for the external medium (\ref{Zeq}), these equations completely specify the dynamics of the system on the Ott-Antonsen manifold.

To look for locked solutions, we plug in the functional form $\bar{z}=re^{i\Omega t}$ into (\ref{Zeq}).  This implies that 
\be
Z(t)=\frac{\rho D r}{i(\Omega+\w_0)+\rho D+J}e^{i\Omega t}.
\label{relZz}
\ee
Inserting the solution for $Z(t)$ back into Eq. (\ref{zbarott}) gives a complex equation for $r$ and the rotation frequency $\Omega$,
\begin{widetext}
\bea
(i\Omega+\Delta)r+\frac{ r\rho D^2}{2}\left[r^2\frac{1}{\rho D+J-i(\Omega+\w_0)}-\frac{1}{\rho D+J+i(\Omega+\w_0)}\right]=0\label{bigeq}
\label{LSlor}
\eea
\end{widetext}
Locked solutions are solutions to this equation with $r> 0$. Notice, that even though the incoherent state, $r=0$, is always a solution, it can be stable or unstable, and the stability boundary is given by the phase boundary (\ref{incolor}). 

We now look for other locked solutions, i.e. those with $r>0$. To do so, we divide (\ref{LSlor}) out by $r$,  and solve for $\Omega$ as a function of $q=r^2$ to get
\be
 \Omega (q)=-\omega_0+\rho D \sqrt{\frac{D}{2\Delta}(1-q)-1}.
\label{Omegaeq}
 \ee
 Substituting the solution back into (\ref{LSlor}) yields a \emph{cubic equation} for $q=r^2$. In general, this expression becomes unwieldy and must be solved numerically. However, we can still draw many conclusions. Physical solutions must have real values of $\Omega$ and real, positive values for $q>0$. Since the end result is a cubic equation for $q$, there can be 0,1,2, or 3 locked solutions for a given parameter range. Each of these solutions can be stable or unstable. We will use these facts to construct a phase diagram in the next section.
 
\subsection{Equations in terms of phase difference}
 
It is useful to rewrite the dynamics of our model in terms of the phase difference, $\psi$, between the order parameter and external medium. Let $Z =Re^{i \Phi}$, $\bar{z}=re^{i\phi}$, and $\psi=\Phi-\phi$. In terms of these variables, we can rewrite the dynamical equations of motion (\ref{zbarott}) and (\ref{Zeq}), as,
\begin{align}
\dot{r}&=-\Delta r+\frac{D}{2}(1-r^2)R\cos\psi  \nonumber \\
\dot{R}&=-(\rho D+J)R+\rho Dr\cos \psi  \label{psiequations} \\
\dot\psi&= -\omega_0-D\frac{R}{2r} \sin\psi\left( 2\rho+r^2+1 \right) \nonumber
\label{psiequations}
\end{align}
These equations fully characterize the dynamics within the Ott-Antonsen manifold. 

For the special case of locked solutions, there is a relationship between the phase difference $\psi$ and the rotation frequency $\Omega$. In particular, by examining the phases of (\ref{relZz}), we have
\be
\cos{\psi}= \frac{\rho D+J}{\sqrt{(\rho D +J)^2+ (\Omega+\omega)^2}}.
\ee
Alternatively, taking the absolute value of (\ref{relZz}) yields a relationship between the magnitude of the order parameter, $r$, and the amplitude of the oscillations in the external medium, $R$,
\be
R=\frac{\rho D}{\sqrt{(\rho D+J)^2+(\Omega+\w_0)^2}} r = \frac{\rho D}{\rho D+J} r \cos{\psi}.
\label{Rrrel}
\ee
Notice that this is consistent with (\ref{psiequations}) since $\dot{R}=0$ for locked solutions.

\section{Phase Diagram and Bistability}
\label{sec:phasediagram}

\subsection{Constructing phase diagram from dynamical equations} 
In this section, we discuss the phase diagram for our model. The dynamics of the external medium leads to a much richer phase diagram than that for the Kuramoto model with direct coupling. For brevity, we focus our discussion on the case where $g(\omega)$ is a Lorentzian distribution. We have numerically verified that a similar phase diagram occurs for other distributions. Furthermore, to reduce the number of parameter, we concentrate on the case where $\Delta=0.1$ and $J=0$. Note that, in contrast to directly coupled oscillators, the solutions exhibit non-trivial $\Delta$ dependence because $\Delta$ must be compared to the mean frequency $\omega_0$. On the other hand, the behavior for $J \neq 0$ is expected to be qualitatively similar to the $J=0$ case because the steady-state solutions in eq. (\ref{bigeq}) depend on $\rho$, $D$, and $J$ only through the two control parameters $\rho D +J$ and $\rho D^2$. 

We can use the results from the previous sections to derive the phase diagram in the $(\rho$-$D)$ plane for various values of $\omega_0$ (see Fig. \ref{fig1}). The red line is the incoherence stability boundary given by (\ref{incolor}). The incoherent solution is stable to the left of this line and unstable to the right. Recall, that in addition to the incoherent solution, there can exist 0,1,2, or 3 locked solutions which can be stable or unstable. These locked solutions are given by the roots of cubic equation for $q=r^2$  resulting from substituting (\ref{Omegaeq}) into (\ref{LSlor}). Furthermore, notice that the only way that the number of locked solutions can change is when one of the roots of the corresponding cubic equation changes sign or if new real ones appear. In order for the former to occur, the solution corresponding to the root changing sign should collide with the incoherent solution $r=0$ and change its stability. Thus, this can occur only at the incoherence boundary given by the red line.  This implies transitions at the incoherence boundary are continuous. On the other hand, new real solutions appear when the discriminant of the corresponding cubic equation equals zero. Note that the phase boundary defined by discriminant equalling is the only place where discontinuous phase transitions can occur.  That is, real positive solutions can come into existence here at non-zero amplitude.  This occurs, for instance, in the transition from Region M1 to B, in the middle panel of Fig. (\ref{fig1}).  These lines are plotted in blue in the phase diagrams.   The number and type of stable equilibria must be identical within each of the regions carved out by the red and blue lines, allowing for a straight forward construction of the phase diagram. 

To check these arguments, we ran numerical simulations of $N=1000$ oscillators with frequencies drawn from a Lorentzian distribution and measured the order parameter after decay of initial transients. We initialized the system with random phases and chose the initial magnitude of $R$ to be either zero or a finite value to probe the stability of the incoherent or partially locked state, respectively. The results were in excellent agreement with the phase diagram in Fig. \ref{fig1}. We also studied a Gaussian $g(\omega)$ distribution and similarly found that the locked solutions exist within the low-dimensional manifold and the system exhibits qualitatively similar behavior to that found for a Lorentzian distribution. 

\subsection{Phase diagram and bistability}
 
We now discuss the phase diagram in greater detail. The top panel in 
Figure \ref{fig1} shows that the typical phase diagram for small $\omega_0$. The phase diagram is similar to that for a directly-coupled Kuramoto model. There are two phases, an incoherence phase and locked phase, with a density-dependent critical coupling $D$ that marks the transition between the two phases. As usual, the order parameter $r$ is zero for the incoherent phase and close to one for the locked solution.

When $\omega_0$ is increased, shown in the middle panel of Fig. \ref{fig1},  a bistable region between incoherence and coherence appears at low densities.  This region results from the subtle interplay between the ``inertia'' of the external medium and the amplitude of the order parameter. The influence of the oscillators  on the external medium is proportional to the density and the amplitude $r$ of the order parameter; see (\ref{psiequations}). Thus, if the oscillators are incoherent  they cannot entrain the media. For large $r$ the opposite is true, giving rise to the bistable region.  At higher densities the phase diagram is topologically similar to the ordinary Kuramoto model with a direct transition between incoherence and coherence.

\begin{figure}[t]
\includegraphics[width=8cm]{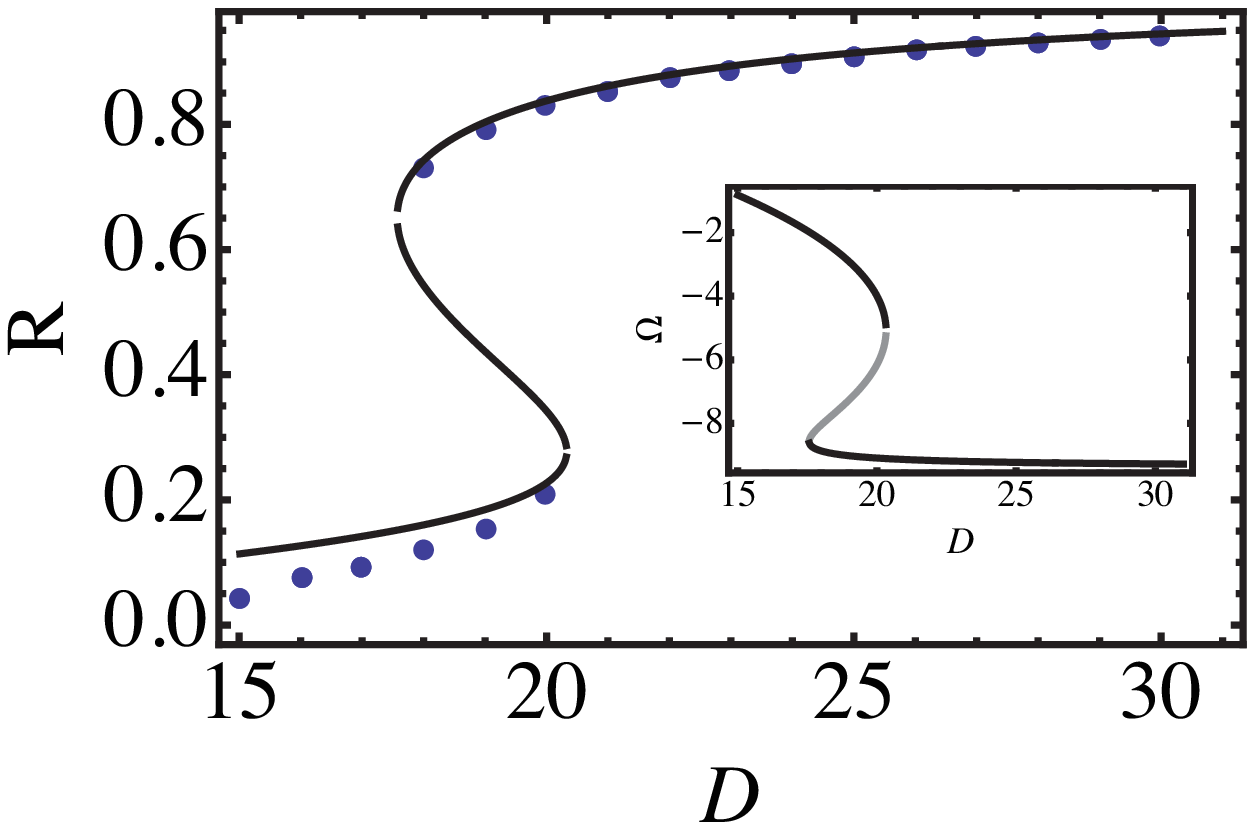}
\includegraphics[width=8cm]{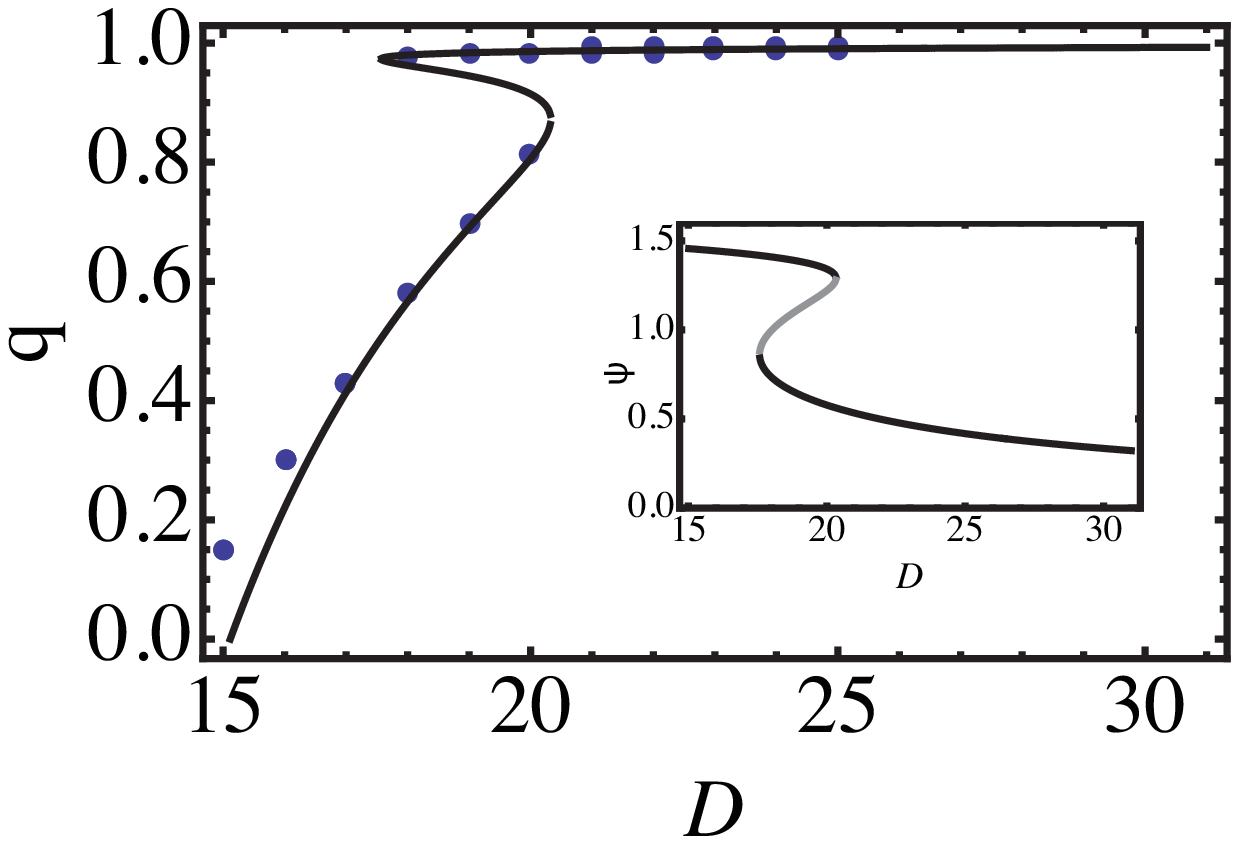}
\caption{Amplitude, $R$, of the external medium (top) and magnitude of the order parameter, $r$, (bottom) for non-trivial locked solutions as a function of coupling $D$, for $\omega_0=10$, and $\Delta=0.1$. The black lines represent stable solutions and gray lines unstable solutions. Notice the bistable region corresponding to region $\tilde{B}$ in  Fig. \ref{fig1}.  Insets show the locked oscillation frequency in the rotating frame (top) and the phase difference, $\psi$, between the medium and order parameter, $z$.}
 \label{fig2}
 \end{figure}

The phase diagram develops more features as $\omega_0$ increased further (bottom of Fig. \ref{fig1}). In addition to all of the behaviors discussed above, there also exists a region of bistabiity between two different locked solutions: one where the amplitude of the external medium oscillations is small and another where the amplitude is large. In both cases, the amplitude of the order parameter $r$ is close to one. The amplitude $R$ and the phase difference between the order parameter and external medium, $\psi$, for these locked solutions can be calculated directly using the aforemetioned cubic equation and are shown in Fig. \ref{fig2}. Notice that the external medium oscillates out of phase with the order parameter for the low amplitude locked solution. Finally, we note that the size of this bistable region increase as $\omega_0$ is increased further.

  \subsection{The low amplitude locked solution}
 
 A novel aspect of coupling oscillators through an external medium is the existence of the aforementioned  locked solution where the order parameter amplitude $r$ is nearly unity but the amplitude of the oscillations in the external medium $R$ is small. Interestingly, this locked solution always  appears together with the usual locked solutions where both $r$ and $R$ are close to one. Figure \ref{fig3} shows numerical simulations  confirming the analytic predictions from the Ott-Antonsen ansatz. This new type of locked solution is possible because the effective coupling between oscillators becomes small in (\ref{Zeq1a}) when $R$ is small even when $D$ is large, allowing oscillators to rotate near their natural frequency. Furthermore, (\ref{Rrrel}) implies that the order parameter and external medium will oscillate at nearly $\pi/2$ out of phase for such a locked state. This can also be seen in the inset of Fig. \ref{fig2}.  

 \section{Relation to Other Systems}
  \label{sec:relmodel}
 
 The model studied here  is closely related to other variants of the Kuramoto model. For $\omega_0=0$, the steady-state dynamics of our mode is identical to that of the Millenium Brigdge Problem  \cite{gc2005crowd, eckhardt2007modeling}. Somewhat more surprisingly, in the $\rho \rightarrow 0$ limit, the steady-state dynamics of our model can be mapped onto the dynamics of the Kuramoto model with a bimodal frequency distribution \cite{martens2009exact}  for the special case when the dynamics of the latter model are restricted to lie on the Ott-Antonsen manifold. We discuss both of these mappings below.
 
\subsection{Millenium Bridge Problem} 
To understand wobbly behavior of the millenium bridge, the authors of  \cite{gc2005crowd}  represented the dynamics of pedestrians walking on a bridge using a simple mathematical model  in which the bridge is modeled as a  driven harmonic oscillator. Pedestrians are also modeled as oscillators that try to phase lock with the bridge. We now show that the steady-state, locked solutions of the class of systems to which the Millenium Bridge belongs can be mapped onto the solutions of our model. 

In the Millenium Bridge Problem, the dynamics of the bridge position, $X$, is governed by the equation
\be
M \frac{d^2X}{dt^2}+B\frac{dX}{dt}+KX=G\sum_{i=1}^N\sin\theta_i\label{bridgeeq}
\ee
with $M, B$, and $K$ the mass, damping, and stiffness of the bridge, respectively. The pedestrian's footsteps, represented by the $\theta_i$. It is also useful to write $X=R\sin\Phi$.
Notice, that for locked solutions where $R=\mbox{cst.}$ and $\Phi=\Omega t$, we can relate the bridge position in the Millenium Bridge Problem to the external medium in the Kuramoto model by noting that  formally we can write $X=\mbox{Im}\{Z\}$. Furthermore, notice that in the non-rotating, lab frame,  isolating the imaginary part of Eq. (\ref{Zeq1}) gives the equation $R\Omega\cos\Phi+(\rho D+J)R\sin\Phi=\frac{\rho D}{N}\sum_j\sin\theta_j$. 

Comparison with the equation resulting from plugging in rotating solution $X=R\sin\Phi$ into (\ref{bridgeeq})  yields a mapping between the steady-state dynamics of the two problems. In particular, it is clear there exist a mapping between parameters  $(\rho,D,J,\omega_0) \to (G, B, K, M)$ for which $R$ and $\Omega$ are preserved. In particular, assuming $R\neq0$ gives the mapping  $\rho D = \alpha G$, $B=1/\alpha$, and $\rho D+J=\alpha(K-M\Omega^2)$, where $\Omega(\rho,D,J,\omega_0)$ is a function of parameters and $\alpha\neq0$ is an arbitrary constant. It is interesting that, due to the freedom in specifying $\alpha$, the bridge can be under-, critically-, or over-damped.

In the original Millenium Bridge Problem, the authors restricted their considerations to case where the natural frequency of the pedestrian and bridge are identical \cite{eckhardt2007modeling}. The rational for this was the ``worst case scenario'' for wobbling of the bridge. Mathematically, this corresponds to choosing $\omega_0=0$ in the equations above. For this choice, the phase diagram is identical to that of the ordinary Kuramoto model with incoherent and coherent phases (see top of Fig. \ref{fig1}) and this was what was  found in \cite{gc2005crowd}. However, as discussed in the last section, the more realistic case where the natural frequency of the pedestrians and bridge differ so that  $\omega_0 \neq 0$, gives rise to a much richer phase diagram seen in this problem. 

 \begin{figure*}[t]
\includegraphics[width=13cm]{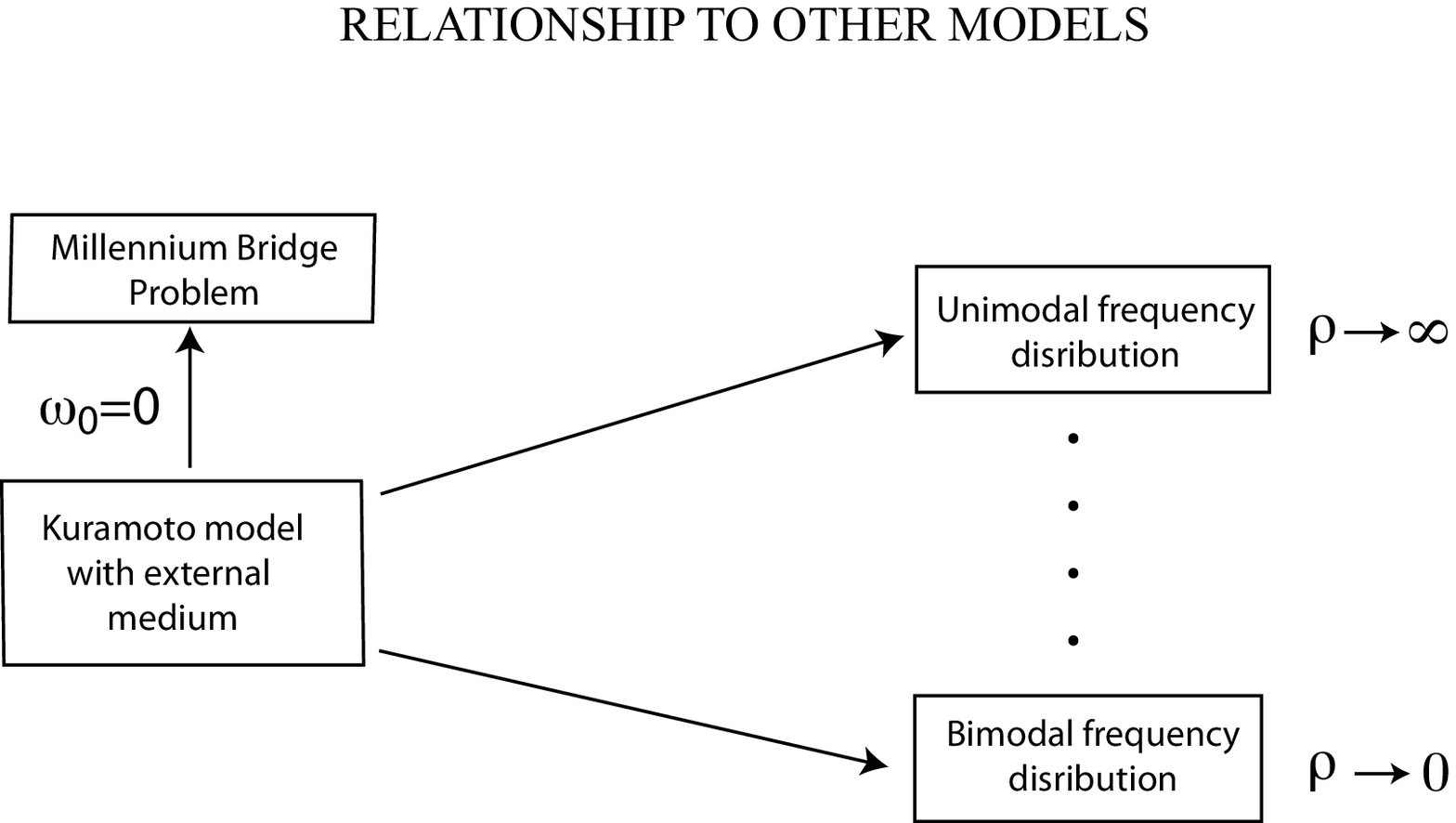}
\caption{Relationship of the Kuramoto model with external medium to other variations of the Kuramoto model. }
 \label{fig4}
 \end{figure*}
 
 \subsection{Bimodal distribution}

Another variant of the Kuramoto model studied recently using the Ott-Antonsen ansatz is the Kuramoto model with a bimodal frequency distribution. The Ott-Antonsen dynamics captured the transition from coherent to incoherent states, including bistability for certain parameters, but failed capture the full dynamics richness of the bimodal model. For example, the Ott-Antonsen dynamics did not allow for standing wave solutions which were seen numerically. Within the Ott-Antonsen manifold, it was argued in \cite{martens2009exact}  that the dynamics of the system is well described by the set of equations for the square of the magnitude of the order parameter $q=r^2$ and a phase difference $\tilde{\psi}$ between the phases of the oscillators locked around $\pm \omega_0$ (see \cite{martens2009exact}) 
\bea
\dot{q} &=& -\tilde{\Delta}q+(q-q^2)[\cos{\tilde{\psi}}+1] ] \nonumber\\
\dot{\tilde{\psi}}&=& \tilde{\omega_0}-(1+q) \sin{\tilde{\psi}},
\label{bimodaleq}
\eea
where the dot indicates a derivative with respect to $\tilde{t}=D/2 t$, $\tilde{\Delta}= (4\Delta)/D$, and $\tilde{\omega_0}= (4 \omega_0)/D$.

We now show that the steady-states of the model considered in this paper are identical to the those derived from the equations above. To do so, it is useful to work with the dynamical equations  (\ref{psiequations}) in terms of $R=|Z|$,  $r=|\bar{z}|$, and the phase difference $\psi$. Furthermore, we concentrate on the case when $J=0$. Since we are interested in steady-states, we can set $\dot{R}=0$ in (\ref{psiequations}) and derive a relationship between $R$ and $r$ given by $ R=r \cos{\psi}$. Substituting this into the remaining two equations gives
\begin{align}
0=\dot{r}&=-\Delta r+\frac{D}{2}(1-r^2)r\cos^2 \psi  \nonumber \\
0=\dot\psi&= -\omega_0-\frac{D}{2} \tan \psi\left( 2\rho+r^2i+1 \right).
\label{Bimodalpsiequations}
\end{align}
Writing $q=r^2$ and taking the limit $\rho \rightarrow 0$ gives
\begin{align}
0=\dot{q}&= -2\Delta q + Dq(1-q)\cos^2\psi \nonumber \\
0=\dot{\psi} &= -\omega_0 -\frac{D}{4} \sin{2\psi}( q+1).
\end{align}
Finally, making the substitution $\tilde{\psi}=2\psi$, $\tilde{t}=D/2 t$, $\tilde{\Delta}=(4\Delta)/D$, and $\tilde{\omega_0}=(4\omega_0)/D$ yields the equations (\ref{bimodaleq}). 

The mapping can also be derived by directly examining equations  (\ref{incolor}) and (\ref{bigeq}). In the limit $\rho \rightarrow 0^=+$, the stability boundary  of the incoherent state (\ref{incolor}) reduces to 
\be
D=2 (\Delta^2+\omega_0^2)/\Delta.
\label{smallrhoinc}
\ee
This is identical to the boundary calculated for directly-coupled oscillators with a bimodal frequency distribution in \cite{martens2009exact} . We can also directly derive an equation for the amplitude of frequency-locked solutions for $\rho=0^+$ by substituting the (\ref{Omegaeq}) into  (\ref{bigeq})add expanding  to lowest order in $\rho$. This procedure yields the equation
\be
\omega_0= \pm \frac{1+q}{1-q}\sqrt{\frac{\Delta}{2}(D-Dq-2\Delta)}.
\label{smallrho}
\ee
As expected, this is identical to the amplitude equation for directly coupled oscillators with a bimodal frequency distribution under the identification $\tilde \omega=4\omega/D$, $\tilde \Delta = 4\Delta/D$. Thus just as in the bimodal case, multiple solutions can arise in our model via a saddle-node bifurcation which occurs when $\partial \omega_0/\partial q = 0$, or equivalently $q=2-\frac{\sqrt{D(D+8\Delta)}}{D}$.  Plugging this expression into (\ref{smallrho}) gives the location of the saddle-node bifurcation with the caveat that when the bifurcation occurs to the left of the incoherence stability boundary, the resulting $q$ is negative and therefore unphysical. By equating this curve with Eq. (\ref{smallrho}), we find new physical solutions with positive $q$ arise when $\omega_0>\Delta/\sqrt{3}$.

Somewhat surprisingly, we have shown that the steady-states of our model in the $\rho, J \rightarrow 0$ limit is identical, with the same parameters, to that of a Kuramoto model with bimodal distribution restricted to the Ott-Antonsen manifold under the identification of the phase difference between the order parameter and external medium in our model with the twice the phase difference between the order parameters of oscillators locked around $\pm \omega_0$ in the bimodal model. This hints that the essential physics of both problems is contained in two coupled oscillators. It would be interesting to explore this further in the future.

\section{Discussion}
In this paper, we have studied the a variation of the Kuramoto model where phase-oscillators are coupled through a common external medium. Such a model is likely to be widely applicable to biological and physical systems where oscillators communicate with each through some chemical signal or physical membrane. An important distinction between the model studied here and the Kuramoto model is that there is an important new control parameter, the density of oscillators in the medium. This allows for interesting new effects such as a density-dependent transition to oscillations which has been dubbed Dynamic Quorum Sensing \cite{de2007dynamical, taylor2009dynamical}. We used the Ott-Antonsen ansatz in combination with numerical simulations to investigate the dynamics of this model. We found that the model had a rich phase diagram with bistability between incoherence and locked solutions, as well as between two different types of locked solutions. In addition, as summarized in Fig. \ref{fig4}, the model is closely related to other variants of the Kuramoto model in various limits.

The underlying reason for the complex phase diagram in our model is the dynamics of the external medium. The external medium has two distinct effects. First, it introduces an effective time-delay for communication between oscillators. This time delay manifests itself mathematically by noting that $\Omega \neq 0$. It was previously shown that introducing a fixed time delay leads to bistability between different locked solutions as is found in this model \cite{yeung1999time}. However, since the time-delay is not fixed in our model, we do not see the hierarchy of locked solutions seen in the direct coupling model with delay. Second, the medium has an ``inertia''  so that the a natural frequency becomes important. In particular, it is precisely when the natural frequency of the external medium is large that the phase diagram of our model differs the most from the ordinary Kuramoto model.

The dynamics of the model presented here are equivalent to the Millenium Bridge problem when $\omega_0=0$. This is unsurprising since it the bridge acts as an effective medium through which walkers communicate. What is somewhat unexpected is the richness in the dynamics that emerges when the resonant frequency of the walkers and bridge differ (i.e. $\omega_0 \neq 0$). More surprisingly, the model with an external medium interpolates between a directly coupled Kuramoto model with unimodal and bimodal distributions as a function of the density $\rho$, with $\rho \rightarrow 0$ corresponding to a bimodal distribution, and $\rho \rightarrow \infty$ a unimodal distribution (see Fig. \ref{fig4}). An important caveat is that this is only true when the dynamics of the Kuramoto model with bimodal distribution is restricted to lie on the Ott-Antonsen manifold. This hints that the underlying the dynamics of many Kuramoto models is captured by the Ott-Antonsen ansatz because the steady-state dynamics reduces to that of a few coupled oscillators. The failure of the ansatz to capture the standing waves in the Kuramoto model with bimodal distribution is likely because this violates the simple picture above. It would be interesting to understand the connection between various Kuramoto-like models further.

Perhaps the most experimentally interesting finding of the paper is the predicted bistability at low densities. It would be interesting to see if this could be observed in an experimental system. The largest obstacle to this is that in systems where oscillators have both a  phase and amplitude, bistability is masked by an oscillator death phase where the amplitude of all oscillators is pulled to zero \cite{schwab2010dynamical}. Thus,  any experimental realization would require that the amplitude of oscillators remain be held fixed and the phase oscillator approximation apply even at strong couplings. Thus, it is unlikely that bistability  exists in experimental setups similar to those used to study dynamical quorum sensing such as the BZ reaction  with catalytic beads \cite{taylor2009dynamical} and quorum-sensing coupled bacteria \cite{danino2010synchronized}. Nonetheless, it will interesting to see if the results here can be experimentally tested.

\section{Acknowledgements}
 We would like to thank Thomas Antonsen, Thomas Gregor,  Troy Mestler, and Javad Noorbakhsh for useful discussions. PM and DS would like to thank the Aspen Center for Physics where part of this work was performed. PM was partially funded by a grant from the National Institute of Health grant number K25GM086909. FUNDING for DS and GP.

\bibliography{refsmain}   

\end{document}